\documentclass[printer]{aa}
\usepackage{amssymb}
\usepackage{graphicx}
\usepackage[utf8]{inputenc}
\usepackage[T1]{fontenc}
\usepackage{amsmath}
\usepackage{natbib}
\usepackage{txfonts}
\usepackage{color}
\bibpunct{(}{)}{;}{a}{}{,} 
\begin{document}

\title{ Origin of  gamma-ray emission in the shell of Cassiopeia A}
\author{L. Saha\inst{1},T. Ergin\inst{2,3}, P. Majumdar\inst{1}, M. Bozkurt\inst{3}\thanks{now at: Tufts University, Physics and Astronomy Department},
          \and E. N. Ercan\inst{3}}

   \institute{Saha Institute of Nuclear Physics, 1/AF, Bidhannagar, Kolkata, 700064, India.\\
              \email{lab.saha@saha.ac.in},
              \and      
         T\"{U}B\.{I}TAK Space Technologies Research Institute, ODT\"{U} Campus, Ankara, 06531, Turkey\\
             \email{tulun.ergin@tubitak.gov.tr}
             \and
             Bo\u{g}azi\c{c}i University, Physics Department, Bebek, Istanbul, 34342, Turkey.
             }





\abstract
{Non-thermal X-ray emission from the shell of Cassiopeia A (Cas A) has 
been an interesting subject of study, as it provides
information about relativistic electrons and their acceleration
mechanisms in the shocks. {\it Chandra} X-ray observatory revealed the detailed spectral and
spatial structure of this SNR in X-rays. The spectral analysis of  {\it Chandra} X-ray data of Cas A shows
unequal flux levels for different regions of the shell, which can be
attributed to different magnetic fields in those regions. Additionally, the GeV gamma-ray emission observed by {\it Large Area Telescope} on board {\it Fermi Gamma Ray Space Telescope} showed that the hadronic processes are dominating in Cas A, a clear signature of acceleration of protons.}
{To locate the origin of gamma rays based on the X-ray data of the shell of Cas A. We also aim to explain the GeV$-$TeV gamma-ray data in the context of both leptonic and hadronic scenario.} 
%
{We modeled the multi-wavelength spectrum of Cas A. We use synchrotron  emission process to explain the observed non-thermal X-ray fluxes from different regions of the shell. These result in estimation of  the model parameters, which are then used to explain TeV gamma-ray emission spectrum. We also use hadronic scenario to explain both GeV and TeV fluxes simultaneously. }
{Based on this analysis, it has been shown that the southern part of the remnant is bright in TeV gamma rays. We also show that, leptonic model alone cannot explain the GeV$-$TeV data. Therefore, we need to invoke a hadronic model to explain the observed GeV$-$TeV fluxes. We found that although pure hadronic model is able to explain the GeV$-$TeV data, lepto-hadronic model provides the best fit to the data.}
{}
 
\authorrunning{L. Saha et al}
\titlerunning{Origin of  gamma-ray emission in the shell of Cassiopeia A}
\maketitle

\section{Introduction}\label{sec:intro}
Cassiopeia A (Cas A) is a historically well-known shell type supernova
remnant (SNR) observed in almost all wavebands, e.g. radio
\citep{baars1977, anderson1995, vinyaikin2007, helmboldt2009}, optical
\citep{reed1995}, IR \citep{smith2009, delaney2010}, and X-rays \citep{allen1997, hwang2004, helder2008, maeda2009}.
 Cas A has been observed in TeV gamma rays by HEGRA \citep{aharonian2001}, MAGIC \citep{albert2007}, and VERITAS \citep{acciari2010} telescopes. Upper limits on the GeV gamma-ray emission was first reported by EGRET \citep{esposito1996}. However, first detection at GeV energies was reported by the \textit{Large Area Telescope} on board the Fermi satellite (Fermi-LAT) \citep{abdo2010}. Brightness of this source in all wavelengths makes it a unique galactic astrophysical source for studying the origin of galactic cosmic rays as well as high-energy phenomena in extreme conditions. The distance to Cas A was estimated to be 3.4 kpc \citep{reed1995}.  

Cas A has a symmetric and unbroken shell structure. Short and clumpy
filaments have also been observed on the outer shell of Cas A \citep{vink2003}. 
Infrared observations \citep{rho2009, rho2012, wallstorm2013} revealed ro-vibrational and high-J rotational CO lines coincident with the reverse shock at the northern, central, and southern parts of Cas A in the form of knots ($\sim$ 0$''\!$.8) of
varying velocity and mass. However, there is no clear evidence, e.g. OH maser
emission, for interaction between dense MC and the shell of Cas A. In X-rays, {\it Chandra} observed the shell of Cas A with high angular accuracy ($\sim$0$''\!$.5). Since the angular resolution is worse for gamma-ray measurements (~360$''$) \citep{albert2007, acciari2010}, Cas A was observed as a point-like source in gamma rays. There is a CCO (compact central object) located very close to the center of Cas A \citep{pavlov2000}. Unlike a typical energetic pulsar \citep{pavlov2004}, a CCO is not capable to produce enough TeV gamma rays for detection. The gamma-ray emission is more likely originating from the shell of the SNR.

Acceleration of cosmic electrons was found at the location of outer shocks  \citep{hughes2000, gotthelf2001, vink2003, bamba2005, patnaude2009} as well as at the reverse shock inside the Cas A \citep{uchiyama2008}. Acceleration of particles to TeV energies was established by HEGRA \citep{aharonian2001}, MAGIC \citep{albert2007} and VERITAS \citep{acciari2010} data. The diffusive shock acceleration mechanism is a well established model
for the acceleration of cosmic-ray particles (both electrons and
protons). \cite{uchiyama2008} showed that the X-ray filaments and knots in the reverse shock are efficient acceleration sites.
The spectral analysis of non-thermal filaments of outermost region of Cas A shell was performed by \cite{araya2010a}. The magnetic fields for different filaments were estimated considering
radiative cooling, advection, and diffusion of  accelerated particles
behind the shock \citep{araya2010a}. Additionally, spectral energy density (SED) of  Cas A at GeV$-$TeV energies for whole remnant was studied by  \cite{atoyan2000} and \cite{araya2010b}, but their study is limited to the whole remnant and doesn't extend to different regions of the shell. Recently, \cite{yuan2013} have reported Fermi-LAT results using 44 months of gamma-ray data, revealing that hadronic emission is dominating in the GeV energy range.  

In this paper, we present the study of five regions, i.e south (S),
south-east (SE), south-west (SW), north-east (NE), north-west (NW), of
the shell of Cas A in X-ray energies, which show different levels of
X-ray fluxes.  We model the X-ray spectra from these regions using
 a leptonic model following \cite{blumenthal1970}. We used the multi-wavelength data i.e. radio data \citep{baars1977},
X-ray data from {\it Chandra} \citep{hwang2004, helder2008},
Fermi-LAT data, MAGIC \citep{albert2007},
and VERITAS \citep{acciari2010} data. We analysed the Fermi-LAT data taken over $\sim$ 60 months of operation. Moreover, we interpreted the multiwavelength  SED
and determined  the region of the shell, which contributes
the most to the total gamma-ray emission from Cas A.



This paper is organized as follows: In Section \ref{sec:DataAnalysis} we show the X-ray analysis results based on \textit{Chandra} observations as well as results from Fermi-LAT data analysis. The multi-wavelength modeling was performed in Section \ref{sec:modeling}. In Sections \ref{sec:discussion} we discuss the results in the context of these multiwavelength observations. Finally, we draw our conclusions in Section \ref{sec:conclusion}.

\section{Data Analysis and Results}\label{sec:DataAnalysis}
\subsection{X-Rays}
We analyzed the Chandra X-ray data of Cas A from April 2004. The data set has an exposure of 166720 seconds at the center position of RA(J2000) = 23$^h$ 23$^m$ 26$^s\!$.70, Dec(J2000) = 58$^{\circ}\!$ 49$'$ 03$''\!$.0 (Obs. id: 4638) \citep{hwang2004}. The X-ray analysis was done for the selected filament regions of S, SE, SW, NE, and NW of the shell in the energy range of 0.7$-$8.0 keV \citep{bozkurt2013, tulun2013}.

\begin{figure}[h]
\centering
\includegraphics[width=0.5\textwidth]{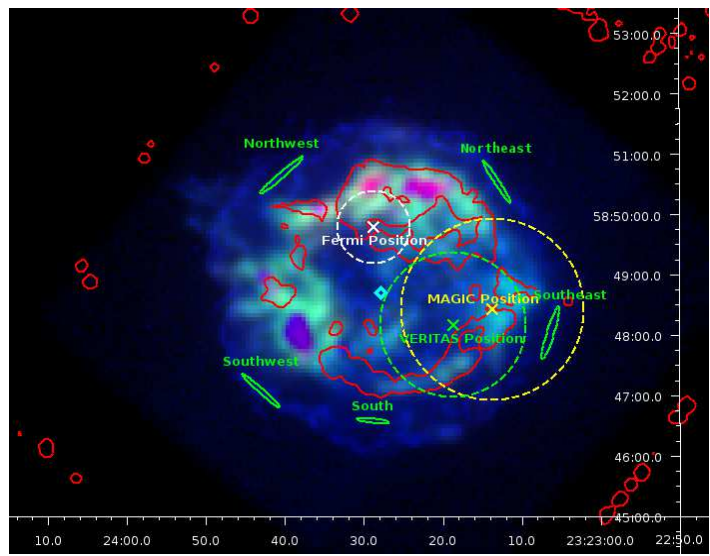}\caption{\small{Multi-color image (Dec vs. RA in J2000) of Cas A produced using Chandra X-ray data. The red, green, and blue color hues represent the energy ranges of [0.7, 1.0], [1.0, 3.5], and [3.5, 8.0] keV, respectively. The red and green hues are smoothed in linear color scale, while the blue hues are shown in logarithmic scale to enhance the view of the smallest number of X-ray counts existing in the outer shell. The green ellipses represent the S, SW, SE, NW, and NE of the shell. The green and yellow crosses and dashed circles correspond to
the VERITAS and MAGIC locations and approximated location error circles. The white cross and dashed circle are for the GeV gamma-ray emission best-fit location and the location error circle from the analysis in this paper. The CCO location is shown with a cyan open diamond.  The  red contours represent the derived CO data from Spitzer-IRAC starting from a value of 0.4 MJy/sr and higher.}}
\label{fig:CasA}
\end{figure}

Figure \ref{fig:CasA} shows the Chandra X-ray image of Cas A, where the blue tones are the highest energy counts (3.5$-$8 keV), while red and green tones are the lower energy ranges of 0.7$-$1.0 and 1.0$-$3.5 keV, respectively. The selected regions contain filaments dominated by non-thermal emission \citep{yamazaki2003,bamba2005, araya2010a, araya2012} which mostly shine in X-ray energies between 3.5 and 8 keV. The locations (with green, yellow, and white crosses) and location errors (with green, yellow, and white dashed circles) of the TeV and GeV gamma-ray emissions as measured by VERITAS, MAGIC, and Fermi-LAT, respectively, are also shown on Figure \ref{fig:CasA}.  CO data derived from Spitzer-IRAC with a starting value of 0.4 MJy/sr and higher is represented by red contours. TeV locations found by VERITAS and MAGIC are more towards the east and south-east of the shell, while the GeV location of Fermi-LAT is towards the inner northern part  of the remnant.  However, the point-spread function of a point-like source of all three detectors is bigger in comparison to the radio size of the shell (5$'$). Therefore, it is not certain, in which part of the shell the GeV and TeV gamma-ray emissions dominate.

\begin{figure}[t]
\centering
\includegraphics[width=0.5\textwidth]{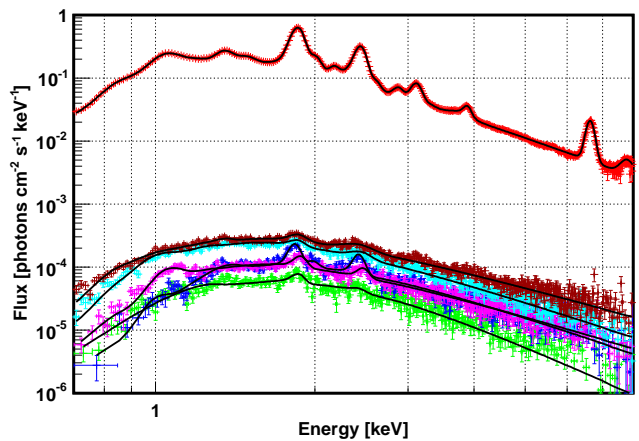}
\caption{X-ray spectra of the five different shell regions and  the whole remnant: (a) green for S, (b) blue for SE, (c) magenta  for SW, (d) cyan for NE, (e) brown for NW, and (f) red for the whole remnant. Corresponding best-fitted X-ray spectra are  shown by black lines.} 
\label{fig:casa_Xray}
\end{figure}

The X-ray spectrum of each selected filament region was first fit with {\it xspec}  power-law by adding {\it wabs} additive model corresponding to photoelectric absorption of hydrogen. Then the spectra were fit to emission lines of iron, silicon, sulphur  using {\it gaussian} components. For each region, we obtained the following fit parameters: spectral index and flux normalization. These two parameters are used for calculating the flux of each region.  The X-ray fluxes of the S, SE, SW, NE, and NW regions are shown in  Fig. \ref{fig:casa_Xray} with data points in green, blue, magenta, cyan, and brown, respectively. The corresponding best-fitted observed  spectra are shown by black lines. We also analyzed the whole remnant in a similar way as that for the selected filament regions. The observed X-ray spectrum corresponding to the whole remnant is represented by the red data points along with the best-fitted spectrum by black line in Fig. \ref{fig:casa_Xray}.

\subsection{Gamma Rays}
We analyzed the GeV gamma-ray data of Fermi-LAT using the standard Fermi Science Tools (FST) package\footnote{http://fermi.gsfc.nasa.gov/ssc/data/analysis/documentation/}  and the alternative package called {\it pointlike} \citep{kerr2011, lande2012}, which is based on {\it gtlike} (FST-v9r27p1). The results of the analysis performed using the standard FST were published in \cite{bozkurt2013}. For the {\it pointlike} analysis, the data used was from 2008-08-04 to 2013-06-26 ($\sim 60$ months) for energies between 200 MeV and 300 GeV. The gamma-ray events were selected from a circular region with radius of 12$^{\circ}\!$ centered at RA(J2000) = 23$^h$ 23$^m$ 25$^s\!$.92 and Dec(J2000) = $+$58$^{\circ}\!$ 48$'$ 00$''\!$.00. Using \emph{gtselect} of FST, the event-type was selected to do galactic point source analysis with Fermi-LAT Pass 7. To prevent event confusion caused by the bright gamma rays from the Earth's limb, we eliminate the gamma rays having reconstructed zenith angles bigger than 105$^{\circ}\!$. The radius of the analysis region (ROI) was chosen as 2$^{\circ}\!$.0. 

The gamma-ray events in the data were binned in  logarithmic energy steps between 200 MeV and 300 GeV. The spectral properties of the gamma-ray emission were studied by comparing the observation with models of possible sources in the ROI. Predictions were made by convolving the spatial distribution and spectrum of the source models with the instrument response function (IRF) and with the exposure of the observation. In the analysis we used the IRF version P7SOURCE$_{-}\!$V6. 

The model of the analysis region contains the diffuse background sources and all point-like sources from the 2nd Fermi-LAT catalog located at a distance equal to or smaller than 1$^{\circ}\!$.8 away from the center of the ROI. The standard background model has two components: diffuse galactic emission (\emph{gal$_{-}$2yearp7v6$_{-}$v0.fits}) and isotropic component (\emph{iso$_{-}$p7v6source.txt}). The background and source modeling was done using the {\it gtlike} and the best set of spectral parameters of the fit were calculated by varying the parameters until the maximum likelihood was maximized. 

The detection of the source is given approximately as the square root
of the test statistics (TS), where larger TS values indicate that the
maximum likelihood value for a model without an additional source (the
null hypothesis) is incorrect. Cas A was detected with a significance
of $\sim$ 37$\sigma$ at the best-fit location of RA =
350$^{\circ}\!$.87 $\pm$ 0$^{\circ}\!$.01 and Dec = 58$^{\circ}\!$.83
$\pm$ 0$^{\circ}\!$.01. The spectrum of Cas  A was modeled by a
power-law function, resulting in a spectral index value of $\Gamma$ =
2.03 $\pm$ 0.02$^{stat}$ and a total photon flux of F$_p$ = (6.17
$\pm$ 0.08$^{stat}$) $\times$ 10$^{-8}$ photons cm$^{-2}$ s$^{-1}$.
The energy flux was found as F$_e$ =  (7.21 $\pm$ 0.05$^{stat}$)
$\times$ 10$^{-11}$ erg cm$^{-2}$ s$^{-1}$ in the energy interval of
0.2 $-$ 300 GeV. The spectral data points of the GeV emission are
shown as  blue triangles in Figure \ref{fig:casa_ic}. In addition, we
have modeled the spectrum with a broken-power-law function and we
obtained the spectral indices to be $\Gamma_1$ = 1.80 $\pm$
0.05$^{stat}$, $\Gamma_2$ = 2.41 $\pm$ 0.06$^{stat}$, and the break
energy at E$_b$ = 4.31 $\pm$ 0.37$^{stat}$ GeV. The total photon and
energy fluxes were found to be F$_p$ = (4.33 $\pm$ 0.14$^{stat}$)
$\times$ 10$^{-8}$ photons cm$^{-2}$ s$^{-1}$ and F$_e$ =  (5.60 $\pm$
0.07$^{stat}$) $\times$ 10$^{-11}$ erg cm$^{-2}$ s$^{-1}$,
respectively. These results are in agreement with the Fermi-LAT
results found by \cite{yuan2013} and \cite{abdo2010}.


\section{Modeling the Spectrum}\label{sec:modeling}

\subsection{Leptonic Model}\label{sec:leptonic_model}
The non-thermal X-ray emission in the selected regions (i.e. S, SE, SW, NE and  NW) can be explained by the synchrotron emission from relativistic electrons in the source.  Since relativistic particle spectra cut-off roughly exponentially based on either acceleration time scale \citep{drury1991} or radiative loss \citep{webb1984}, we considered a relativistic electron distribution following a power-law with an exponential cut-off as shown in Eq. (\ref{eqn:electron_spectrum}):
\begin{eqnarray}
{dN \over d\gamma} = \rm{N_e}~ \gamma^{-\alpha} ~ \exp{\left(-{\gamma \over \gamma_{max}}\right)},
\label{eqn:electron_spectrum}
\end{eqnarray}
where $\mbox{N}_e$ and $\gamma_{\rm max}$ are the constant of proportionality of electron distribution and the Lorentz factor of the cut-off energy of the  electrons, respectively.
The spectrum of the  synchrotron radiation for a power-law distributed electrons can be written as \citep{blumenthal1970}
\begin{eqnarray}
\mbox{F}_{\nu} \propto \mbox{N}_e ~\mbox{B}^{(\alpha+1)/2} \nu ^{-(\alpha+1)/2}
\label{eqn:synchrotron}
\end{eqnarray}
where B is the magnetic field in the emission volume. Eq. (\ref{eqn:synchrotron}) shows that the synchrotron radiation depends  on three  parameters, i.e. $\mbox{N}_e, ~\mbox{B},~ \mbox{and}~ \alpha$.   
 From the observed radio spectrum, $S_{\nu}  \propto \nu^{-0.77}$ \citep{baars1977},  the power-law spectral index, $\alpha$, is estimated  to be 2.54. Hence, the value of $\mbox{F}_{\nu}$ now changes with $\mbox{N}_e$ (which is a measure of the electron density  in the SNR) and magnetic field (B). \cite{atoyan2000} developed a two-zone model, where the magnetic fields from the radio knots (zone one) and from the shell of the remnant (zone two)  were estimated based on the observed radio data. In this two-zone scenario of Cas A, the magnetic field energy densities are different for two different zones, but the density of relativistic electrons in these zones are comparable to each other. Hence, in our model calculations, we used the uniform electron density for all the shell regions. If $\mbox{N}_e$ is fixed, then the level of synchrotron flux will  depend only on the magnetic field. 
\begin{table}[h]
\begin{center}
\caption{The magnetic field parameters for the synchrotron spectra for all selected regions.  }
\label{table_1}
\begin{tabular}{l c}
\hline \hline Region              & Magnetic Field (B) [$\mu$G]   \\ \hline
                     South                 & 250                                        \\ \hline
           Southwest        & 330                                             \\ \hline
           Southeast         & 330                                              \\ \hline
           Northeast         & 410                                             \\ \hline
           Northwest         & 510                                               \\ \hline
\end{tabular}

\end{center}
\end{table}

\begin{figure}[t]
\centering
\includegraphics[width=0.5\textwidth]{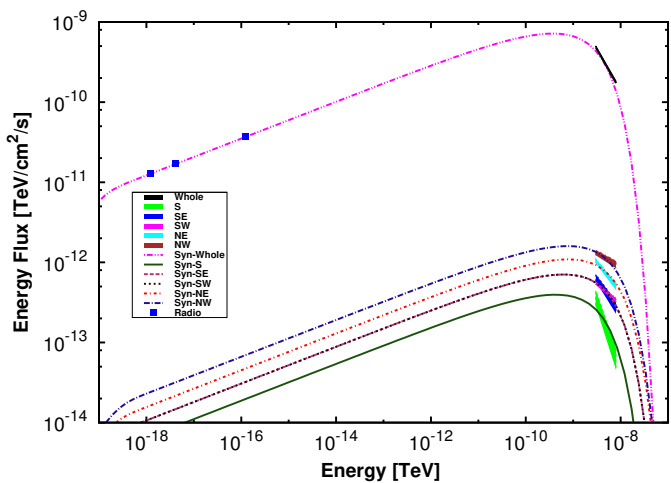}
\caption{Synchrotron  spectra along with the  observed best-fitted X-ray data  for different regions of the shell. The spectral index  describing radio emitting electrons for all the regions is $\alpha$ = 2.54. The estimated magnetic fields for different shell  regions are: 250 $\mu$G for S (solid line); 330 $\mu$G for SE (dashed line) and SW (dotted line); 410 $\mu$G for NE (dot-dashed line); 510 $\mu$G for NW (long dash-dotted line) and 250 $\mu$G for the whole remnant (double dot dashed line). The best-fitted X-ray fluxes of the S, SE, SW, NE, NW and the whole remnant are shown with green, blue, magenta, cyan, brown and black stripes.}
\label{fig:casa_syn}
\end{figure}

{\it Chandra} observations from different regions of the shell show unequal levels of X-ray fluxes, which are attributed to different magnetic fields in those regions.  These different magnetic fields thus contribute to the gamma-ray fluxes through inverse Compton (IC) and bremsstrahlung processes. To estimate the contribution to gamma rays, we first considered that the whole remnant is uniform in X-ray flux and the southern part of the remnant contains only a fraction of the flux from the whole remnant. In the two-zone model of Cas A \citep{atoyan2000}, the mean magnetic field in the shell region was found to be 300 $\mu$G and the energy content of relativistic electrons in this region  was calculated to be about $10^{48}$ erg.  Here, we assumed somewhat lower magnetic field, i.e. 250 $\mu$G for the S region of the shell and the other model parameters for the synchrotron emission process were estimated from the  fit to the corresponding observed best-fitted X-ray data. Afterwards, we estimated the magnetic fields for all other regions using same electron density and same power-law spectral index. The calculated magnetic field values for all the shell regions are shown in Table \ref{table_1}. Based on the flux upper limit given by SAS-2 and COS B detectors, a lower limit on the magnetic field in the shell of Cas A was estimated to be 80 $\mu$G \citep{cowsik1980} and our estimated magnetic fields  do not violate this lower limit on the magnetic field. The rest of the parameters are fixed for all the regions, which are the following: spectral index, $\alpha$ = 2.54, $\gamma_{max}$ = 3.2 $\times$ 10$^7$, distance = 3.4 kpc. The fitted synchrotron spectra and  the observed best-fitted X-ray spectra  for different regions are shown by lines and stripes, respectively in Fig. \ref{fig:casa_syn}.  

\begin{figure}[t]
\centering
\includegraphics[width=0.5\textwidth]{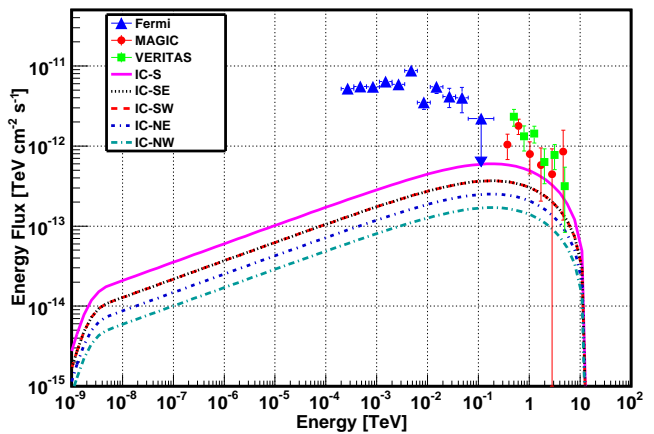}
\caption{IC spectra of the whole remnant  based on parameters related to  different parts of the shell. The spectra for those regions are shown by the following lines: S  by solid line,  SE region by dotted line, SW by dashed line, NE by dot-dashed line, and NW by long dashed-dotted line. The spectra for SE and SW are overlapping. The parameters for radio emitting electrons are $\alpha = 2.54,~ \gamma_{max}~=~3.2 \times 10^7$. }
\label{fig:casa_ic}
\end{figure}

 The inverse Compton (IC) emission spectrum for the whole remnant was estimated using the parameters of the leptonic model obtained for each of the regions in the shell.  IC spectrum  can be estimated in two different ways for each of the regions in the shell. First, the magnetic field for each region was considered to be the  mean magnetic field for the whole remnant. Then parameters for input spectrum were obtained from the fit to the observed radio and X-ray data. Once all the parameters  for leptonic model were estimated, IC  spectrum could be calculated. Secondly, we multiplied the radio synchrotron spectrum for each region with a scale factor, which was estimated by dividing the whole remnant's observed radio or X-ray flux by the corresponding estimated synchrotron flux of this shell region.  
The IC spectra, considering scattering of electrons on far infra-red dust emission at T = 97 K, which dominates over cosmic microwave background photons as seed photons in the emission process \citep{atoyan2000},  are shown in Fig. \ref{fig:casa_ic} for the shell regions.  If the shell region is dominated by strong magnetic field (e.g. 510 $\mu$G), then the IC component of radiation is reduced, which is evident from  Fig. \ref{fig:casa_ic}. It shows that the TeV flux from  the S region of the shell is higher than those from  other regions of Cas A. 

\begin{table}[h]
\begin{center}
\caption{Parameters for Bremsstrahlung process for the S region of the SNR.}
\begin{tabular}{l c }
\hline \hline Parameters                           & Values                                                   \\ \hline
           $\gamma_{max}$                & 3.2 $\times$ 10$^7$                                \\ \hline
           n$_{\rm H}$                                    & 10 cm$^{-3}$                                       \\ \hline
           $\alpha$                                & 2.54                                                        \\ \hline
           Energy ($W_e$)                                      & 4.8 $\times$ 10$^{48}$ erg           \\ \hline
\end{tabular}

\label{table_2}
\end{center}
\end{table}

It is evident from Fig. \ref{fig:casa_ic} that the TeV data at higher energy bins fits better with the IC prediction for the S region among all other regions.  However, the Fermi-LAT spectral data points at GeV energies can not be explained by the IC mechanism alone and it has to be modeled by  an additional component, like the bremsstrahlung process or  the neutral pion decay model. 

\begin{figure}[t]
\centering
\includegraphics[width=0.5\textwidth]{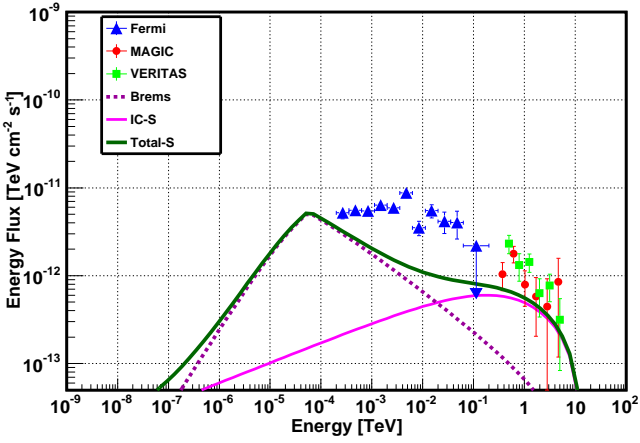}
\caption{Gamma-ray spectrum for Cas A. IC (solid line) and bremsstrahlung (dashed line) spectra are estimated for the whole remnant when parameters are based on  S region of the shell. Bremsstrahlung spectrum  is calculated for n$_{\rm H}$ = 10 cm$^{-3}$. The thick solid line corresponds to total contribution to gamma rays from leptons. The parameters for radio emitting electrons are $\alpha = 2.54,~ \gamma_{max}~=~3.2 \times 10^7$. }
\label{fig:casa_ic_brem}
\end{figure}

Therefore, we estimated the contribution of bremsstrahlung process to explain fluxes at both GeV and TeV energies. The parameters corresponding to  the S region  was used  to calculate bremsstrahlung spectrum considering ambient proton density to be n$_{\rm H}$ = 10 cm$^{-3}$ \citep{laming2003}.  The total energy of the electrons was estimated to be $W_e$ =  4.8 $\times$ 10$^{48}$ erg. The parameters used for this emission process is shown in Table \ref{table_2}. Fig. \ref{fig:casa_ic_brem} shows that  the bremsstrahlung process  alone can not explain the GeV and TeV data simultaneously. Since, the bremsstrahlung flux depends linearly  on the ambient proton density, higher values of ambient proton density  can increase the  GeV$-$TeV fluxes to observed fluxes at these energies. Using mass of supernova ejecta, M$_{ejecta}$ = 2M$_{\odot}$ \citep{willingale2003, laming2003}, where M$_{\odot}$ is the solar mass, the effective gas density was found to be n$_{eff}$ $\simeq$ 32 cm$^{-3}$ \citep{abdo2010}.  It is also not possible to explain GeV$-$TeV data with this density of ambient gas.  Moreover, Fig. \ref{fig:casa_ic_brem} shows that the shape of the observed GeV spectrum near 1 GeV is different from that of  bremsstrahlung spectrum, which rises as it goes from GeV to lower energies.  


\begin{figure}[t]
\centering
\includegraphics[width=0.5\textwidth]{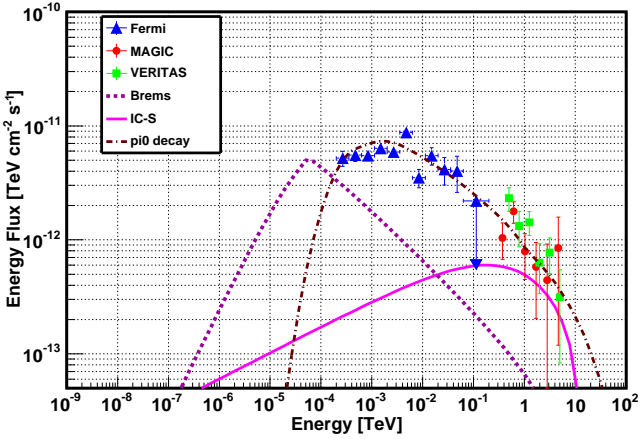}
\caption{Same as Fig. \ref{fig:casa_ic_brem}. Only $\pi^{0}$ decay spectrum for the power-law distributed proton  spectra  (long dash-dotted line) is included. The parameters used to get the $\pi^{0}$ decay spectrum are shown in Table \ref{table_3} by parameter Set-I. The estimated total energy of the protons, $W_p$ = 5.7 $\times$ 10$^{49}$ erg. }
\label{fig:casa_pi0}
\end{figure}

\begin{figure}[t]
\centering
\includegraphics[width=0.5\textwidth]{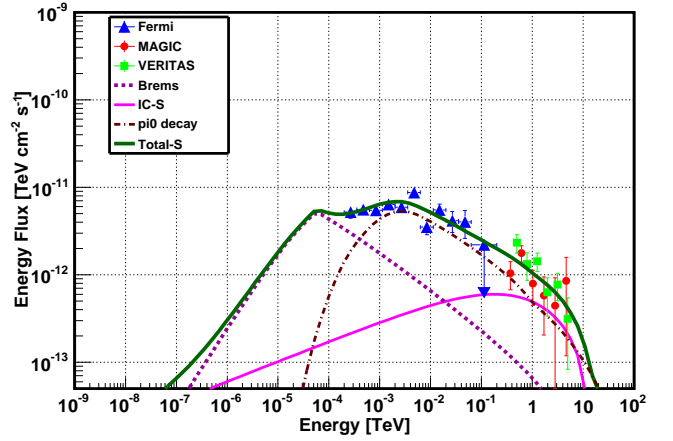}
\caption{Gamma-ray spectrum (thick solid line) for Cas A for combining both leptonic and hadronic contribution to the whole remnant data. Parameters for leptonic model spectra (IC: solid line and bremsstrahlung: dashed line) correspond to S region of  the remnant. Parameters for hadronic model are shown in Table \ref{table_3} by  Set-II (long dash-dotted line).   }
\label{fig:proton_gamma_spec_param2}
\end{figure}

\subsection{Hadronic Model}
Since, leptonic model is not able to account for the observed gamma-ray emission at GeV energies, we need to invoke hadronic scenario to explain observed GeV fluxes. The gamma-ray flux resulting from the neutral pion ($\pi^{0}$) decay of accelerated protons was estimated considering ambient proton density to be 10 cm$^{-3}$. The accelerated protons  were considered to follow a broken power-law spectrum with an exponential cut-off  as shown in Eq. (\ref{eqn:proton_spectrum}).
\begin{eqnarray}
{dN \over dE_p} & = & N_1 ~ E_p^{-\rho}~ \mbox{for}~ E_p^{min}~ \leq E_p ~<E_p^{break} \nonumber \\
        & = & N_2  ~ E_p^{-\beta}~ \exp{\left(-{E \over E_p^{max}}\right)}~ \mbox{for}~ E_p^{break} \leq~ E_p ,
        \label{eqn:proton_spectrum}
\end{eqnarray}
where $N_1$ and $N_2$ are two normalization constants and $\alpha$ and $\beta$ are spectral indices before and after the break at $E_p^{break}$.  Fig. \ref{fig:casa_pi0} shows the contribution of gamma-ray flux from the $\pi^{0}$ decay calculated  following \cite{kelner2006}. The gamma-ray spectrum was fitted within the observed GeV$-$TeV energy range (see Fig. \ref{fig:casa_pi0}) and the corresponding best fit parameters are shown in Set-I of Table \ref{table_3}.   The total energy of the protons in hadronic model was estimated to be $W_p$ = 5.7 $\times$ 10$^{49} {\left(10~ \rm{cm^{-3}}/ n_H \right)}$ erg. We would like to note that we are considering gamma-ray spectrum for the whole remnant, because the angular resolution for the current generation gamma-ray instruments are not comparable to the X-ray instruments.

It has been already mentioned in the Section \ref{sec:leptonic_model} that the leptonic model can only account for the TeV fluxes at the highest energy bins, whereas the observed GeV fluxes can be explained by hadronic model as shown in Fig. \ref{fig:casa_pi0}. So, to get a complete understanding of the spectrum at GeV$-$TeV energies, we have to estimate the combined spectrum resulting from both leptonic and hadronic model (hereafter, lepto-hadronic model).  Since, the parameters for the leptonic model is fixed by observed radio and X-ray fluxes, the resulting parameters for the  hadronic model for the combined GeV$-$TeV spectrum will be  different from the parameters listed in Table \ref{table_3} (see Set-I).  The best fit parameters for the lepto-hadronic model are given in Set-II of Table \ref{table_3} and it shows that the corresponding $\chi^2$ value is less than that of purely hadronic model. The corresponding spectrum is shown in Fig. \ref{fig:proton_gamma_spec_param2}. The maximum energy of protons ($E_p^{max}$) was fixed to 100 TeV for both pure hadronic and lepto-hadronic model. The total energy of the  charged particles for the lepto-hadronic model is $W_e +W_p = 3.4 \times 10^{49}$ erg, which gives a conversion efficiency of supernova explosion energy to be less than 2\%, consistent with  that value calculated by \cite{yuan2013}. It is  to be mentioned that although the GeV data corresponds to the best fit location as shown in Fig. \ref{fig:CasA} using white dashed circle, it doesn't mean gamma rays are not being emitted from other regions of  Cas A. Hence, there is no harm in combing GeV$-$TeV data to get best fitted emission model. 

\begin{table}[h]
\begin{center}
\caption{Parameters for gamma-ray production through decay of neutral pions.}
\begin{tabular}{ l c c}
\hline \hline  
Parameters &  Set-I  &  Set-II \\
&           (hadronic) & (hadronic+leptonic)\\
\hline
$\rho$   & $2.05 \pm 0.05$ & $1.26\pm 0.2$ \\ \hline
$\beta$   & $2.36 \pm 0.02$ & $2.44\pm0.03$  \\ \hline
$E_p^{max}$ (TeV) & 100          & 100         \\ \hline
$E_p^{break}$ (GeV)& 17       & 17  \\ \hline
Energy ($W_p$) (erg)  & 5.7 $\times$ 10$^{49}$  & 2.97 $\times$ 10$^{49}$  \\ \hline
$\chi^2/dof$  & 2.5                      & 1.8 \\ \hline 
\end{tabular}

\label{table_3}
\end{center}
\end{table}


\section{ Discussion}\label{sec:discussion}
The aim of this paper is to locate the region of the shell of Cas A which is brightest in  gamma rays  and to interpret the observed fluxes at GeV$-$TeV energies in the context of both leptonic and hadronic models.
From the different levels of X-ray flux in several shell regions, we found that the magnetic fields are different in those regions. Since, IC flux becomes less significant with higher magnetic field, NW region of the shell of the remnant  is less significant in producing gamma-ray fluxes through IC process among all other shell regions. On the other hand, S region of the remnant becomes a significant region for production of IC fluxes due to  the lower magnetic field in this region. Although we have considered same spectral index for radio emitting electrons, different choices of spectral indices will not change the overall conclusion. 

In addition to that, we see from Fig. \ref{fig:casa_ic_brem} that while bremsstrahlung  process is  unable to explain the GeV$-$TeV data for ambient gas density 10 $\rm cm^{-3}$, IC fluxes can explain the  data only at TeV energies. Moreover, the bremsstrahlung process was also unable to explain the observed fluxes properly with higher values of ambient gas density  ($\sim$ 32 $\rm cm^{-3}$). Therefore, we conclude that the leptonic scenario is insufficient  to explain the observed GeV and TeV gamma-ray fluxes simultaneously. Hence, we need to invoke hadronic contribution to account for the observed gamma-ray fluxes. The gamma-ray spectrum resulting from decay of  $\pi^0$s in Fig. \ref{fig:casa_pi0} shows that it can alone explain the GeV and TeV fluxes with a $\chi^2$ value of 2.5. Since, we have already seen that the leptonic scenario can contribute to TeV energies, we can not ignore this completely. So, we estimated the total contribution from both leptonic and hadronic model to explain the data.  Fig. \ref{fig:proton_gamma_spec_param2} shows that  the gamma-ray spectrum due to decay of $\pi^0$  along with leptonic model is  able to explain the GeV$-$TeV gamma rays for the ambient gas density of 10 $\rm cm^{-3}$. Moreover, the best fit $\chi^2$ value for this case is less than that of the case of purely hadronic model  as shown in Table \ref{table_3}.  Although increasing the effective density of the ambient gas to higher values (than the estimated average density) may help the bremsstrahlung model to reach the level of GeV$-$TeV data, the gamma-ray fluxes due to the $\pi^{0}$ decay of accelerated protons also increase. So, the $\pi^{0}$ decay process will no longer be able  explain the GeV$-$TeV data, unless the total energy budget of the protons is reduced. That in turn indicates a lower conversion efficiency of the explosion energy of Cas A into accelerating protons. The GeV flux falling below 1 GeV  is considered to be a clear indication for the $\pi^{0}$ decay origin of gamma-ray emission. Very recently \cite{yuan2013} reported that their Fermi-LAT data analysis of Cas A  resulted in the gamma-ray emission to be hadronic in nature. A higher density of ambient gas may establish the fact of having higher potential for producing gamma rays through $\pi^0-$decay process for the S region. Nevertheless, total contribution to TeV energies due to both leptonic and hadronic models cannot be ignored. 


According to \cite{vink2003}, the magnetic field at the forward shock is within the range of  80$-$160 $\mu$G whereas the mean magnetic field value in  the shell of Cas A was estimated to be $\sim$ 300 $\mu$G by \cite{atoyan2000, parizot2006}. Although, \cite{abdo2010} showed that leptonic model for the magnetic field of 120 $\mu$G can broadly explain the observed GeV$-$TeV fluxes, the corresponding spectrum from leptonic model didn't fit well. With lower value of the magnetic field (i.e. $<$ 120 $\mu$G), the TeV fluxes will be overestimated  by IC emission spectrum. If we  consider the magnetic field  for the S region of the shell to be  about 120 $\mu$G,  lepto-hadronic model will have to be sufficiently modified to explain the total fluxes. The total fluxes from lepto-hadronic model will  exceed the observed values and the corresponding $\chi^2$ value for the best fit parameters will become large. Also, this overestimated fluxes can not be compensated by lowering the contribution from hadronic model. Hence,  we need to consider higher values of magnetic field ($\sim$ 250 $\mu$G), so that total spectrum from lepto-hadronic model can explain the data sufficiently better than both a pure leptonic and a pure hadronic model.

The most interesting result is the relatively low total accelerated particle energy (of the order of 2\% conversion efficiency) combined with the high magnetic fields estimated. This amplification of magnetic field either be related through magneto-hydrodynamic waves generated by cosmic rays \citep{bell2001, bell2004} or could result from the effect of turbulent density fluctuations on the propagating hydrodynamic shock waves, which has been observed through two-dimensional magneto-hydrodynamic numerical simulations \citep{giacalone2007}. Low conversion efficiency of cosmic rays suggests that the cosmic ray streaming energy may not be sufficient enough to be transferred to the magnetic fields resulting magnetic amplification. Hence the magnetic field amplification in the down-stream of shocks due to presence of turbulence  could be favourable in this particular remnant. 

It is to be mentioned that the differences in the X-ray flux levels for  different regions of Cas A, can be attributed to the different densities of the  injection of the electrons. 
But for  different densities, there may not be any difference in gamma-ray fluxes from different regions of the shell. However, this needs a detailed investigation of the density profile of the relativistic electrons in this source. 


\section{Conclusion}\label{sec:conclusion}
We found that the gamma-ray emission from the S region of the shell through IC process has the highest flux value and this predicted flux matches better to the TeV data in comparison to flux predictions from other regions of the shell. The second best fitting IC prediction with the TeV data is from the SE and SW, and then NE region. We also found that the leptonic model alone is unable to explain observed GeV fluxes for any regions of the shell. But, the GeV and TeV gamma-ray data fits reasonably well to the hadronic model, which is independent of the selected regions on the shell. 

If the SNR would be perfectly symmetric in shape, we would expect that the fluxes of each region of the shell should be equal. But apparently, the shell's emission is not homogeneously distributed.  The reason for the variations in X-ray and gamma-ray fluxes can be due to different amounts of particles or variations in the magnetic field at different regions of the SNR's shell. Also the molecular environment might be different at different sides of the shell. Future gamma-ray instruments with far better angular resolution (e.g. CTA) will be required to understand the spectral and spatial structure of the remnant in gamma rays. 

\begin{acknowledgements}
We would like to thank Prof. Jeonghee Rho for providing us CO data from Spitzer-IRAC. T. E. acknowledges support from the Scientific and Technological
Research Council of Turkey (T\"{U}B\.{I}TAK) through the B\.{I}DEB-2232
fellowship program. This work was supported by Bo\u{g}azi\c{c}i
University. E. N. E. thanks to Bo\~gazi\c{c}i University BAP support
through a project of code 5052P.
\end{acknowledgements}

\bibliographystyle{aa}
\bibliography{}
\end{document}